\documentclass[preprint,aps,prl,nofootinbib]{revtex4-1}
\usepackage[utf8]{inputenc}
\DeclareUnicodeCharacter{FB01}{fi}

\usepackage{amsmath}    
\usepackage{graphicx}   
\usepackage{verbatim}   
\usepackage{color}      
\usepackage{epstopdf}  
\usepackage{upgreek} 	
\usepackage[hidelinks]{hyperref}   
\usepackage{textcomp} 
\usepackage{siunitx}
\usepackage{mathtools}
\DeclarePairedDelimiter\abs{\lvert}{\rvert}

\newcommand{\Mumax}{MuMax$^{3}$ }
\newcommand{\figref}[1]{Fig.~\ref{#1}}

\begin{document}

\title{Magnetic chirality controlled by the interlayer exchange interaction}
\author{Mariëlle J. Meijer}
\email{m.j.meijer@tue.nl}
\affiliation{Department of Applied Physics, Eindhoven University of Technology, P.O. Box 513, 5600 MB Eindhoven, the Netherlands}
\author{Juriaan Lucassen}
\affiliation{Department of Applied Physics, Eindhoven University of Technology, P.O. Box 513, 5600 MB Eindhoven, the Netherlands}
\author{Oleg Kurnosikov}
\affiliation{Department of Applied Physics, Eindhoven University of Technology, P.O. Box 513, 5600 MB Eindhoven, the Netherlands}
\author{Fabian Kloodt-Twesten}
\author{Robert Fr\"{o}mter}
\affiliation{Universität Hamburg, Center for Hybrid Nanostructures, Luruper Chaussee 149, 22761 Hamburg, Germany}
\author{Rembert A. Duine}
\affiliation{Department of Applied Physics, Eindhoven University of Technology, P.O. Box 513, 5600 MB Eindhoven, the Netherlands}
\affiliation{Institute for Theoretical Physics, Utrecht University, Leuvenlaan 4, 3584 CE Utrecht, the Netherlands}
\author{Henk J.M. Swagten}
\affiliation{Department of Applied Physics, Eindhoven University of Technology, P.O. Box 513, 5600 MB Eindhoven, the Netherlands}
\author{Bert Koopmans}
\affiliation{Department of Applied Physics, Eindhoven University of Technology, P.O. Box 513, 5600 MB Eindhoven, the Netherlands}
\author{Reinoud Lavrijsen}
\affiliation{Department of Applied Physics, Eindhoven University of Technology, P.O. Box 513, 5600 MB Eindhoven, the Netherlands}

\date{\today}
\begin{abstract}
Chiral magnetism, wherein there is a preferred sense of rotation of the magnetization, has become a key aspect for future spintronic applications. It determines the chiral nature of magnetic textures, such as skyrmions, domain walls or spin spirals, and a specific magnetic chirality is often required for spintronic applications. Current research focuses on identifying and controlling the interactions that define the magnetic chirality. The influence of the interfacial Dzyaloshinskii-Moriya interaction (iDMI) and, recently, the dipolar interactions have previously been reported. Here, we experimentally demonstrate that an indirect interlayer exchange interaction can be used as an additional tool to effectively manipulate the magnetic chirality. We image the chirality of magnetic domain walls in a coupled bilayer system using scanning electron microscopy with polarization analysis (SEMPA). Upon increasing the interlayer exchange coupling, we induce a transition of the magnetic chirality from clockwise rotating N\'eel walls to degenerate Bloch-N\'eel domain walls and we confirm our findings with micromagnetic simulations. In multi-layered systems relevant for skyrmion research a uniform magnetic chirality across the magnetic layers is often desired. Additional simulations show that this can be achieved for reduced iDMI values when exploiting the interlayer exchange interaction. This work opens up new ways to control and tailor the magnetic chirality by the interlayer exchange interaction. 

\end{abstract} 
\maketitle
Magnetic chirality corresponds to a preferred sense of rotation of the magnetization and understanding this chirality has become of great importance for new spintronic applications \cite{Fert2017,doi:10.1063/1.5048972,Nagaosa2013}. These applications rely on the chiral nature of magnetic textures, like skyrmions or domain walls. In future magnetic memory devices, for instance the racetrack memory \cite{Fert2013}, a controlled displacement of skyrmions or domain walls is of utmost importance for a reliable operation and a key requisite for this is a uniform magnetic chirality of the magnetic textures \cite{thiaville2012dynamics,Emori2013,Ryu2013,sampaio2013nucleation,Woo2016}. Current research focuses on identifying and controlling the interactions that define the magnetic chirality. Understanding the underlying mechanisms will allow one to tailor the magnetic chirality \cite{Hrabec2017} for spintronic applications.  

The most promising interaction that allows for the control of the magnetic chirality is the interfacial Dzyaloshinskii-Moriya interaction (iDMI), which has been studied extensively in the past years in magnetic thin films \cite{Han2016,yang2018controlling,PhysRevLett.120.157204,PhysRevB.100.100402}. This interaction is an anti-symmetric exchange interaction and originates from a broken symmetry at the interface of a ferromagnet and heavy metal \cite{PhysRevLett.87.037203,Bode2007,PhysRevB.78.140403}. The strength and sign of the iDMI depends on the specific material combination at an interface and the iDMI energetically favors either a clockwise (CW) or counterclockwise (CCW) rotation of the magnetization. This allows for the stabilization of magnetic textures, like skyrmions, N\'eel domain walls or spin spirals, with a uniform magnetic chirality \cite{Fert2017,Bode2007}.

Very recently, it was recognized that dipolar fields also influence the magnetic chirality \cite{PhysRevB.98.104402,PhysRevApplied.10.064042,Legrandeaat0415,2019arXiv190103652F,Dovzhenko2018,PhysRevLett.123.157201}. Although the effects of the dipolar interaction were already known for a long time \cite{HubertDomains,malozemoff1979magnetic,kambersky1995domains,labrune1999,Tekielak2011}, their impact on magnetic thin-film systems hosting an iDMI was only recently observed when stacking several magnetic thin-films on top of each other. These magnetic multi-layers are commonly used to stabilize skyrmions at room temperature \cite{Luchaire_skyrmion,Woo2016} and the increased magnetic volume leads to stronger dipolar fields. As a result, the dipolar field emitted from out-of-plane magnetized domains can influence the in-plane magnetization, which results in a non-uniform magnetic chirality across the magnetic multi-layers. Various models \cite{PhysRevB.98.104402,PhysRevApplied.10.064042,Legrandeaat0415} and first experiments \cite{Dovzhenko2018} show that this behavior can be generalized and has a profound impact on many magnetic textures such as skyrmions and derived entities. For most spintronic applications the stabilization of a uniform magnetic chirality across a multi-layered system is desired, \cite{Legrandeaat0415,PhysRevB.98.104402} which can be achieved by implementing a strong iDMI to overcome the dipolar interaction. Generating a strong iDMI is not always achievable, however, and severely constrains the design of the multi-layered system. 

\begin{figure}
\centering
\includegraphics{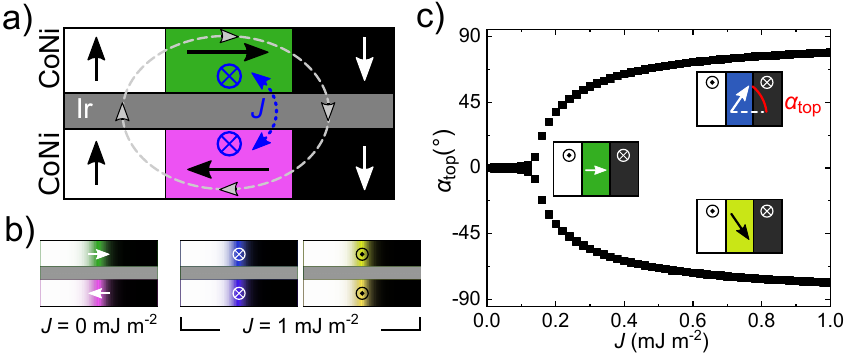}
\caption{\label{fig:figure1} a) Elementary model of two magnetic CoNi layers separated by an Ir spacer layer in the absence of an iDMI (side view). The up and down domains are indicated by the white and black areas respectively, and they generate dipolar fields (gray dashed lines). The in-plane magnetization direction of the domain wall aligns with the dipolar field (along the arrow), resulting in CW/CCW N\'eel walls in the top/bottom magnetic layer. A ferromagnetic RKKY interaction (blue dashed line) rotates the in-plane magnetization towards Bloch walls (blue arrows). b) Micromagnetic simulation results for $J=0$~\si{mJ.m^{-2}} and the two degenerate cases for $J=1$~\si{mJ.m^{-2}}. The in-plane magnetization direction is indicated by the arrow and the same color indications as in a) are used. c) Angle $\alpha_\mathrm{top}$ as a function of RKKY coupling strength $J$ obtained from micromagnetic simulations. $\alpha_\mathrm{top}$ defines the angle between the in-plane magnetization direction and the horizontal of the top magnetic layer (see inset). The insets show a top view of the magnetization direction for $J=0$~\si{mJ.m^{-2}} (left inset) and $J=0.4$~\si{mJ.m^{-2}} (right insets for the degenerate case).}
\end{figure}

In this article, we demonstrate an alternative approach to control the magnetic chirality utilizing the effect of an indirect interlayer exchange interaction \cite{STILES1999322,PhysRevB.52.411,HELLWIG200713,doi:10.1063/1.4922726,PhysRevLett.116.177202}, namely the conventional Ruderman–Kittel–Kasuya–Yosida (RKKY)\footnote{The recently discovered asymmetric exchange component of the RKKY interaction is not present in this work \cite{2018arXiv181001801F,2018arXiv180901080H}.} interaction \cite{PhysRevLett.67.3598,Luo_1998,PhysRevB.52.411}. First, we determine the influence of the ferromagnetic RKKY interaction on the magnetic chirality by imaging the domain wall magnetization in a bilayer system with negligible iDMI using scanning electron microscopy with polarization analysis (SEMPA) \cite{Oepen2005,UNGURIS2001167,doi:10.1063/1.3534832,PhysRevB.96.060410}. In the absence of the RKKY interaction the dipolar fields cause a non-uniform magnetic chirality in the bilayer system, and this results in the formation of CW N\'eel walls in the top magnetic layer and CCW N\'eel walls in the bottom magnetic layer as is schematically depicted in \figref{fig:figure1}a. Upon increasing the ferromagnetic RKKY coupling, the magnetization in the domain walls asymptotically rotates towards non-chiral Bloch walls. In the second part we investigate a multi-layered  system including iDMI typically used for skyrmion research with the help of micromagnetic simulations. We explicitly show that the necessary iDMI values to obtain a uniform magnetic chirality can be reduced by 30\% in the presence of a strong ferromagnetic RKKY interaction. Utilizing the RKKY interaction therefore opens up new ways to tune and control the chirality of magnetic textures on a layer-by-layer basis. 

Before we show our dedicated sample design and illustrate the obtained experimental results, we would like to first address the basic physical principles of how the dipolar fields and in particular the ferromagnetic RKKY influence the magnetic chirality in the absence of an iDMI. We concentrate on an elementary model consisting of two magnetic CoNi layers RKKY-coupled via an Ir spacer layer as depicted in \figref{fig:figure1}a, which mimics the experimental situation. Both layers exhibit a perpendicular magnetic anisotropy and the up and down domains (white and black areas, respectively) of the magnetic bilayer generate dipolar fields as indicated by the grey dashed line. The in-plane magnetization direction inside the domain walls aligns with the dipolar fields as depicted by the arrows in the green and pink area and this leads to the formation of a CW N\'eel wall in the top magnetic layer and a CCW N\'eel wall in the bottom magnetic layer. By coupling the magnetic layers ferromagnetically (dashed blue line) this anti-parallel alignment of the magnetization in the domain wall can be counteracted, resulting in the stabilization of degenerate Bloch walls pointing either into the paper (as indicated by the blue arrows) or out of the paper (not shown).

We confirm the validity of this intuitive picture using \Mumax \cite{Vansteenkiste2014,De_Clercq_2017} micromagnetic simulations, with the simulation conditions specified in supplementary SI. On the left side of \figref{fig:figure1}b the result in the absence of a ferromagnetic RKKY coupling ($J=0$~\si{mJ.m^{-2}}) is depicted and the formation of a CW/CCW N\'eel wall in the top/bottom magnetic layer is found, respectively, as expected from the dipolar interaction. Introducing a ferromagnetic RKKY coupling ($J=1$~\si{mJ.m^{-2}}) leads to the formation of two energetically degenerate Bloch walls, as depicted on the right side of \figref{fig:figure1}b, where the in-plane magnetization direction of both magnetic layers points either into the paper (blue area) or out of the paper (yellow area). We therefore find, that a uniform magnetization profile across the magnetic layers in a bilayer system can be achieved due to the presence of a ferromagnetic RKKY interaction. A preferred chirality is not present, however, since two kinds of Bloch domain walls can be stabilized. In \figref{fig:figure1}c we study the transition between N\'eel and Bloch walls as a function of $J$ in more detail. Here, we focus on the domain wall formation in the top magnetic layer and the angle $\alpha_\mathrm{top}$ describes the in-plane magnetization direction of the domain wall as depicted schematically in the top right inset. We find that $\alpha_\mathrm{top}=0^\circ$ for $J=0$~\si{mJ.m^{-2}} (see left inset) and $\alpha_\mathrm{top}$ asymptotically approaches the formation of Bloch walls ($\alpha_\mathrm{top}=\pm90^\circ$) for large $J$. For intermediate values of $J$ degenerate Bloch-N\'eel domain walls are formed and this is schematically depicted in the insets on the right side for a value of $J=0.4$~\si{mJ.m^{-2}}. The micromagnetic results indicate that the ferromagnetic RKKY interaction influences the magnetic chirality and that the strength of the interaction determines the in-plane magnetization direction of the domain walls.  

\begin{figure*}
\centering
\includegraphics{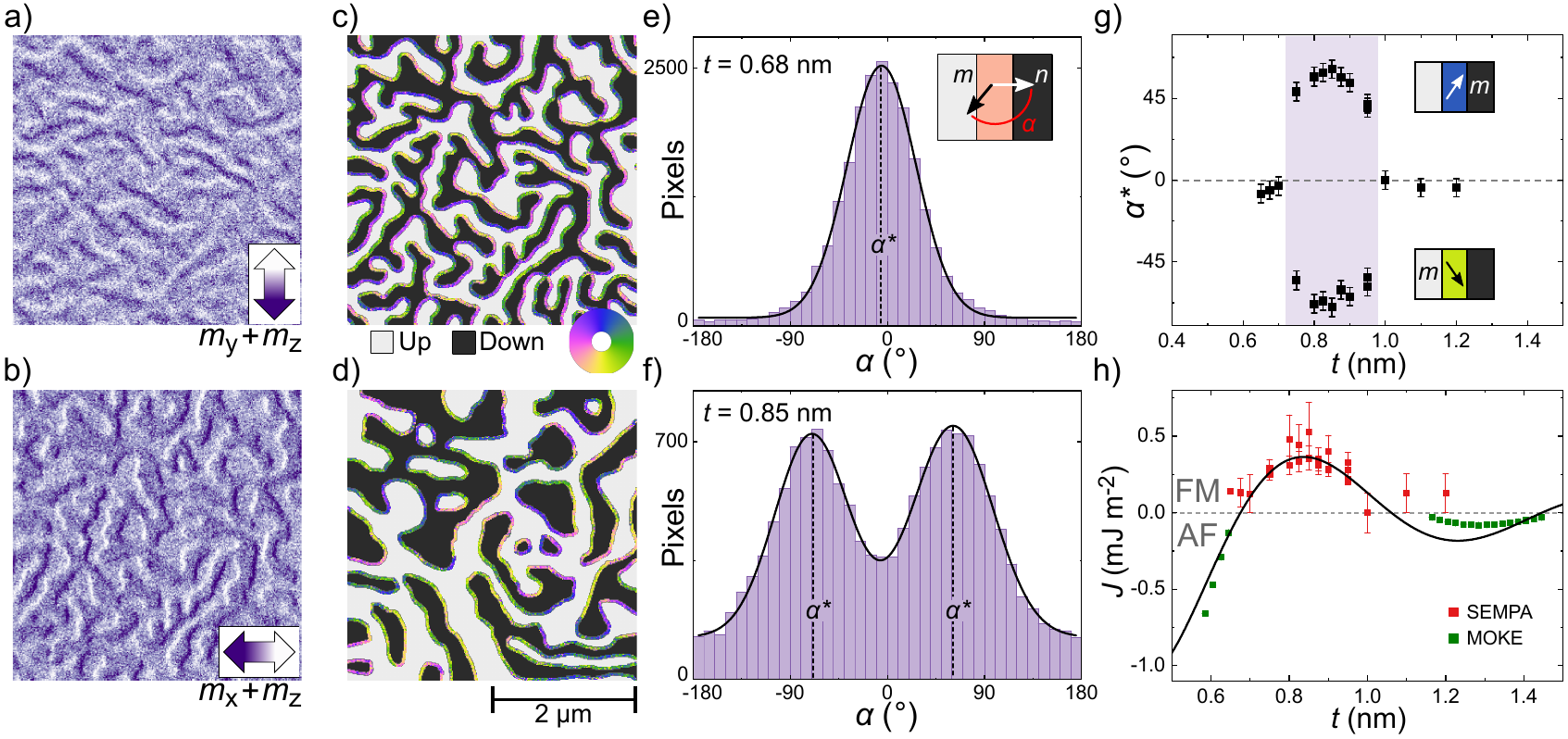}
\caption{\label{fig:figure2} a,b) SEMPA images of the top magnetic layer at an Ir thickness of $t=0.68$~\si{nm}. Image a) shows $m_\mathrm{y}$ + out-of-plane contrast ($m_\mathrm{z}$) and image b) $m_\mathrm{x}$ + out-of-plane contrast ($m_\mathrm{z}$) for the same area. The in-plane magnetic contrast direction is indicated by the arrow in the bottom right corner. c) Composite image constructed from the SEMPA images in a) and b), with the in-plane magnetization indicated by the color wheel and the out-of-plane contrast by the white and black areas (up and down magnetization, respectively). d) Composite image at an Ir thickness of $t=0.85$~\si{nm}. The same scale bar is used for all images. e,f) Histograms of the angle $\alpha$ for all pixels in the domain walls of image c) and d), respectively. $\alpha$ is defined as the difference between the domain wall normal $n$ and magnetization in the domain wall $m$ (see inset of e), and c) for the color indications). The solid line is a (double) Gaussian fit with the maximum(s) at $\alpha^\ast$. g) Maximum(s)  $\alpha^\ast$ of the Gaussian fits of the histograms as a function of Ir thickness $t$. In the purple-shaded area two maximums are found and the insets schematically show the corresponding magnetization texture in this region. The color indications from c) are used. h) RKKY coupling strength $J$ as a function of Ir spacer layer thickness $t$. The MOKE data is extracted from hysteresis loops (see supplementary SIII for details) and the SEMPA data in g) is translated to a value of $J$ with the help of \figref{fig:figure1}c. Both data sets are fitted with the theoretical RKKY function from Ref. \cite{PhysRevB.52.411} (solid black curve).}
\end{figure*}

In the following we experimentally measure this influence by imaging the domain wall chirality in a bilayer system with SEMPA for different RKKY coupling strengths. Here, we map the magnetization profile of specifically the top magnetic layer, due to the high surface sensitivity of SEMPA. From literature it is known that Iridium mediates a strong RKKY interaction that alternates between an antiferromagnetic (AF, $J<0$) and ferromagnetic (FM, $J>0$) coupling as a function of thickness $t$ with a damped sinusoidal behavior \cite{STILES1999322,PhysRevLett.67.3598,Luo_1998,PhysRevB.52.411}. We therefore grew a sample with the following composition: //Ta(3)/Pt(3)/[Co(0.6)/Ni(0.35)]$_\text{x2}$Co(0.2)/Ir($t$) /[Co(0.6)/Ni(0.35)]$_\text{x2}$ (thicknesses in parentheses in nm), where the Ir thickness is wedged from $t=0.5-1.5$~\si{nm}, providing access to both the ferromagnetic and antiferromagnetic coupling regime of the RKKY interaction (see Methods for more details). 

In \figref{fig:figure2}a and b SEMPA images of the top magnetic layer are depicted at an Ir thickness of $t=0.68$~\si{nm}. The images were measured simultaneously and show the same area. \figref{fig:figure2}a displays the $m_\mathrm{y}$ magnetization contrast and \figref{fig:figure2}b the $m_\mathrm{x}$ magnetization contrast, as indicated by the arrows in the bottom right corner. In both images a slight out-of-plane contrast is also visible (see Methods), where the lighter areas correspond to up domains and the darker areas to down domains. The domains are framed by dark or light bands, which correspond to the in-plane component of the magnetization in the domain wall. The combined information of the SEMPA images is depicted in the composite image of \figref{fig:figure2}c using a procedure described elsewhere \cite{PhysRevLett.123.157201}. Here, the out-of-plane contrast is indicated by the white and dark areas (up and down, respectively), and the in-plane magnetization direction in the domain wall is indicated by the color wheel. From the composite image we find that the domain walls on the left side of an up domain are pink, which is equivalent to the magnetization pointing towards the left, whereas the magnetization on the right side of an up domain points towards the right, illustrated by the green color. This indicates that the magnetization in the domain wall always points from an up domain towards a down domain and reveals the presence of CW N\'eel walls. We can investigate this more thoroughly by defining an angle $\alpha$, which is the difference between the domain wall normal $n$ and the magnetization direction $m$, as depicted in the inset of \figref{fig:figure2}e. Assigning this angle $\alpha$ to every pixel in the domain wall results in the histogram shown in \figref{fig:figure2}e. Around $\alpha=0^\circ$ a peak in the histogram is observed that corresponds to the formation of CW N\'eel walls. The histogram is fitted with a Gaussian curve that models the underlying statistics of the individual pixels \cite{PhysRevLett.123.157201} and allows one to extract the peak position $\alpha^\ast$. For an Ir thickness of $t=0.85$~\si{nm} the same procedure results in the composite image shown in \figref{fig:figure2}d and the corresponding histogram is depicted in \figref{fig:figure2}f. In the histogram two distinct peaks are observed and their position is extracted with a double Gaussian fit giving $\alpha^\ast=-70^\circ\pm5^\circ$ and $\alpha^\ast=61^\circ\pm5^\circ$. The two types of domain walls that are stabilized in \figref{fig:figure2}d,f are neither CW N\'eel walls ($\alpha^\ast=0^\circ$) nor Bloch walls ($\alpha^\ast=\pm90^\circ$), but show rather an intermediate Bloch-N\'eel texture as is schematically depicted in the insets of \figref{fig:figure1}c. Additional measurements for different Ir thicknesses can be found in supplementary SII. 

The extracted $\alpha^\ast$ is plotted as a function of Ir thickness $t$ in \figref{fig:figure2}g. CW N\'eel walls are formed for $t<0.75$~\si{nm} and $t>1.0$~\si{nm} and the two degenerate Bloch-N\'eel walls are present for intermediate Ir thicknesses in the purple-shaded area. Within this shaded area an increase in $\abs{\alpha^\ast}$ is observed for increasing $t$, until a maximum is reached at $t=0.85$~\si{nm}, whereafter $\abs{\alpha^\ast}$ decreases again. According to the findings presented in \figref{fig:figure1}c, the CW N\'eel walls are stabilized by the dipolar interaction. The formation of the degenerate Bloch-N\'eel walls in the purple-shaded area can then be explained by the interplay between the dipolar interaction and ferromagnetic RKKY interaction. Since $\abs{\alpha^\ast}$ scales with $J$, we find the expected increase and decrease of the ferromagnetic RKKY coupling as a function of the Ir thickness.

To further substantiate that the interlayer exchange interaction is the dominant mechanism that stabilizes the degenerate Bloch-N\'eel walls, we study the expected oscillatory behavior of the RKKY interaction in more detail. Therefore, we combine the information on the coupling strength in both the ferromagnetic and antiferromagnetic region as a function of the Ir layer thickness $t$ and this is plotted in \figref{fig:figure2}h. The data in the ferromagnetic region is plotted in red and obtained via the SEMPA measurements discussed previously, where the angle $\alpha^\ast$ from \figref{fig:figure2}g is converted to a coupling strength $J$ using the micromagnetic simulations presented in \figref{fig:figure1}c. Information on the coupling strength in the antiferromagnetic RKKY region can be obtained from the switching fields in the antiferromagnetic hysteresis loops measured by the magneto-optical Kerr effect (MOKE), as is explained in more detail in supplementary SIII. In \figref{fig:figure2}h the coupling values in the antiferromagnetic region are plotted in green. When we combine both data sets we clearly observe the oscillatory behavior of the RKKY coupling $J$ as a function of the Ir thickness $t$ and the data is fitted with the theoretically predicted RKKY behavior \cite{PhysRevB.52.411} (solid black curve). The theory describes the periodic behavior well and a maximum ferromagnetic coupling of approximately $0.4$~\si{mJ.m^{-2}} is obtained at $t=0.85$~\si{nm}. Both the extracted period of the oscillation as well as the RKKY coupling strength are in agreement with values found in literature \cite{PhysRevMaterials.3.041401,PhysRevLett.67.3598,Luo_1998}. 

So far, we have seen experimentally and from micromagnetic simulations that in the absence of an iDMI the RKKY interaction influences the magnetic chirality induced by the dipolar interaction. Moreover, the simulations of \figref{fig:figure1}b indicate that the difference between the in-plane magnetization directions of the domain walls across the bilayer system is reduced due to the ferromagnetic RKKY coupling. For a strong coupling almost identical magnetic textures are stabilized in the top and bottom magnetic layer. A uniform magnetic chirality can not be obtained by the ferromagnetic RKKY interaction alone, however, due to the degeneracy of the Bloch-(N\'eel) walls. Adding an iDMI can lift this degeneracy and a uniform chirality across the magnetic layers can be achieved. 

\begin{figure}
\centering
\includegraphics{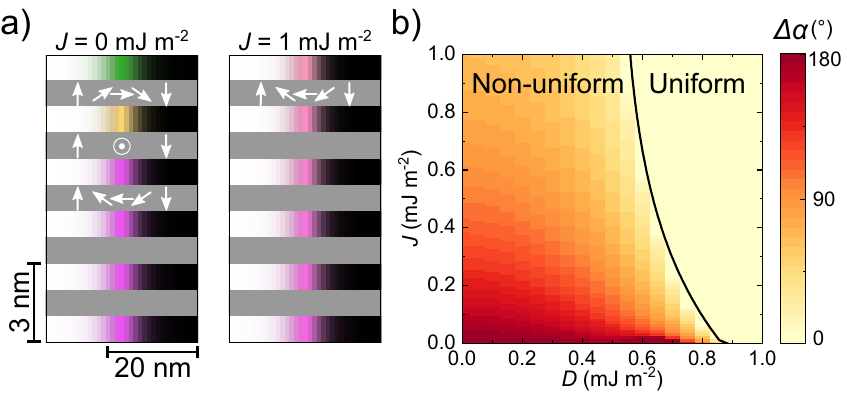}
\caption{\label{fig:figure3} a) Micromagnetic simulations of a multi-layer stack containing 6 magnetic layers with $D=0.5$~\si{mJ.m^{-2}}. The up and down domains are indicated by the white and black areas respectively, and the colors in the domain walls show the in-plane magnetization direction according to the color wheel of \figref{fig:figure2}c. The arrows in the grey spacer layer indicate the magnetization direction inside the magnetic layer above. In the left image $J=0$~\si{mJ.m^{-2}} and in the right image $J=1$~\si{mJ.m^{-2}}. b) Phase diagram of the angle $\Delta\alpha$ as a function of $J$ and $D$. $\Delta\alpha$ is defined as the difference between $\alpha$ in the bottom and top magnetic layer. For $\Delta\alpha=0^\circ$ there is a uniform chirality throughout the multi-layer stack. The black line indicates the transition between a non-uniform and a uniform magnetization. }
\end{figure} 

In the following we examine the necessary conditions to obtain a uniform magnetic chirality across the magnetic multi-layers, when the dipolar interaction, RKKY interaction and iDMI are present. We study this with micromagnetic simulations in a multi-layered system typically hosting chiral magnetic textures like skyrmions. The investigated multi-layered stack consists of 6 repeats with alternating magnetic and spacer layers of $1$~\si{nm} (details of the simulations and the dependence on saturation magnetization $M_S$ and effective anisotropy $K_\mathrm{eff}$ values can be found in the supplementary SI and SIV). In \figref{fig:figure3}a the magnetic textures obtained for two RKKY strengths are depicted with an iDMI of $D=0.5$~\si{mJ.m^{-2}}. In the left image $J=0$~\si{mJ.m^{-2}} and a non-uniform magnetization texture is observed across the magnetic layers. The bottom layers form CCW N\'eel walls, favored by the positive $D$, but the iDMI is not strong enough to counteract the dipolar interaction. This results in the formation of a Bloch wall and CW N\'eel in the top two layers. We define the uniformity of the chirality in the multi-layered system by subtracting the $\alpha$ values from the bottom and top magnetic layer and this results in $\Delta\alpha=180^\circ$ for the case of $J=0$~\si{mJ.m^{-2}}. In the right image of \figref{fig:figure3}a $J=1$~\si{mJ.m^{-2}} and an approximately uniform chirality in all the magnetic layers is achieved with $\Delta\alpha=7^\circ$. In \figref{fig:figure3}b $\Delta\alpha$ is plotted for a range of $D$ and $J$ values. Two regions are indicated where the chirality is either non-uniform ($\Delta\alpha\neq0^\circ$) or uniform ($\Delta\alpha=0^\circ$). Without an RKKY interaction ($J=0$~\si{mJ.m^{-2}}) an iDMI value of at least $D=0.9$~\si{mJ.m^{-2}} is needed to stabilize a uniform chirality and this corresponds to the critical iDMI value of the system. Interestingly, this critical iDMI value can be reduced by approximately 30\% when an RKKY interaction of $J=1$~\si{mJ.m^{-2}} is present, as can be seen from the transition line in \figref{fig:figure3}b. In practice these $D$ and $J$ values can be achieved in magnetic multi-layers by optimizing the thicknesses and materials of the magnetic and non-magnetic spacer layer \cite{PhysRevLett.67.3598,Han2016}, which makes it possible to stabilize magnetic textures with a uniform chirality in a wider variety of multi-layered systems than previously assumed. 

Finally, an additional effect of the interlayer exchange coupling on the magnetic texture becomes apparent when we compare the images of \figref{fig:figure2}c and d. The average domain size grows as the ferromagnetic RKKY interaction increases and this is elaborated in more detail in supplementary SV. Although the influence of the iDMI is not considered yet, the findings suggest that the RKKY interaction might be used to control the size of magnetic domains and possibly even skyrmions.   

To conclude, we have demonstrated that an interlayer exchange interaction influences the magnetic chirality. In a system where dipolar fields are present, the influence of the RKKY interaction manifests itself as a rotation of the magnetization in the top domain wall from a CW N\'eel to a degenerate Bloch-N\'eel wall. We confirm these findings by micromagnetic simulations. Furthermore, micromagnetic simulations predict that the RKKY interaction reduces the iDMI required to obtain a uniform magnetic chirality across a typical multi-layer system for skyrmion research. Making use of the well-known interlayer exchange interaction opens up new ways to tune and control the magnetic chirality in multi-layered systems for spintronic applications. 

This work is part of the research programme of the Foundation for Fundamental Research on Matter (FOM), which is part of the Netherlands Organisation for Scientific Research (NWO). We acknowledge financial support by the DFG within SFB 668. R. A. D. also acknowledges the support of the European Research Council.

\section{Methods}
The sample that is investigated consists of two magnetic CoNi layers coupled via an Ir spacer layer, with the following composition: //Ta(3)/Pt(3)/[Co(0.6)/Ni(0.35)]$_\text{x2}$Co(0.2)/Ir($t$) /[Co(0.6)/Ni(0.35)]$_\text{x2}$ (thicknesses in parentheses in nm). We wedged the Ir layer $t$ between $t=0.5-1.5$~\si{nm} to vary the strength of the RKKY coupling. The sample was grown on a Si substrate with a native oxide layer by DC sputter deposition. The base pressure of the system is $4\times10^{-9}$~\si{mbar} and the Ar pressure during deposition was $1\times10^{-2}$~\si{mbar}. The saturation magnetization $M_\mathrm{S}$ and anisotropy $K$ values can be found in supplementary SI. We optimized the thickness composition of Co and Ni such that the sample shows a perpendicular magnetic anisotropy, but is close to the spin-reorientation transition. This ensures that the as-deposited sample is in a multi-domain state and can be imaged directly as we transfer the sample \textit{in-situ} to the SEMPA setup. With SEMPA we can map the complete in-plane magnetization vector \cite{Oepen2005,UNGURIS2001167,doi:10.1063/1.3534832}. Additionally, we can distinguish up and down domains by tilting the sample stage, which results in a projection of the out-of-plane magnetization on the in-plane measurement axis \cite{PhysRevB.96.060410,PhysRevLett.100.207202,doi:10.1063/1.4998535}. The out-of-plane contrast in the $m_\mathrm{y}$ image is well defined and adjustable by rotation of the sample stage. The out-of-plane contrast in $m_\mathrm{x}$ depends strongly on the sample mounting. From the SEMPA measurements a composite image as depicted in \figref{fig:figure2}c and d can be obtained and the method is described in Ref. \cite{PhysRevLett.123.157201}. This reference additionally shows that the widths of the histograms in \figref{fig:figure2}e and f is mainly the result of Poisson noise in the electron counting \cite{doi:10.1063/1.3534832} and errors in the extraction of the domain wall normal\footnote{Note that in reference \cite{PhysRevLett.123.157201} the domain wall normal is rotated by $180^\circ$ compared to definition found in this article.}.

\bibliography{references}
\end{document}


\title{Supplementary Information: Magnetic chirality controlled by the interlayer exchange interaction}
\author{Mariëlle J. Meijer}
\email{m.j.meijer@tue.nl}
\affiliation{Department of Applied Physics, Eindhoven University of Technology, P.O. Box 513, 5600 MB Eindhoven, the Netherlands}
\author{Juriaan Lucassen}
\affiliation{Department of Applied Physics, Eindhoven University of Technology, P.O. Box 513, 5600 MB Eindhoven, the Netherlands}
\author{Oleg Kurnosikov}
\affiliation{Department of Applied Physics, Eindhoven University of Technology, P.O. Box 513, 5600 MB Eindhoven, the Netherlands}
\author{Fabian Kloodt-Twesten}
\author{Robert Fr\"{o}mter}
\affiliation{Universität Hamburg, Center for Hybrid Nanostructures, Luruper Chaussee 149, 22761 Hamburg, Germany}
\author{Rembert A. Duine}
\affiliation{Department of Applied Physics, Eindhoven University of Technology, P.O. Box 513, 5600 MB Eindhoven, the Netherlands}
\affiliation{Institute for Theoretical Physics, Utrecht University, Leuvenlaan 4, 3584 CE Utrecht, the Netherlands}
\author{Henk J.M. Swagten}
\affiliation{Department of Applied Physics, Eindhoven University of Technology, P.O. Box 513, 5600 MB Eindhoven, the Netherlands}
\author{Bert Koopmans}
\affiliation{Department of Applied Physics, Eindhoven University of Technology, P.O. Box 513, 5600 MB Eindhoven, the Netherlands}
\author{Reinoud Lavrijsen}
\affiliation{Department of Applied Physics, Eindhoven University of Technology, P.O. Box 513, 5600 MB Eindhoven, the Netherlands}
\date{\today}
\maketitle

\section{Details on micromagnetic simulations and sample parameters} \label{mumax}
For the micromagnetic simulations we used \Mumax{}\cite{Vansteenkiste2014}. To produce Fig. 1b and c of the main paper the following settings for the simulations were used. The cell sizes were $1\times 8 \times 0.1$~nm (x, y, z) with periodic boundary conditions in the $x$ and $y$ direction equal to 32 repeats. We explicitly introduce a $1.0$~\si{nm} thick spacer layer to account for the Ir and the simulation box in the $x$ and $y$ direction was $256\times32$~nm. We used $M_\mathrm{S}=1.13$~\si{MA.m^{-1}} and $K=1.07$~\si{MJ.m^{-3}} given by the result of the SQUID-VSM measurements on //Ta(3)/Pt(3)/[Co(0.6)/Ni(0.35)]x2/Co(0.2)/Ir(0.95)/[Co(0.6)/Ni(0.35)]x2/Ta(3) using the area method to determine the anisotropy \cite{0034-4885-59-11-002}. In the main paper we took an exchange stiffness $A=12$~\si{pJ.m^{-1}}, which is a slightly lower value then typically used for thin-film Co, due to the added Ni in the magnetic layer \cite{Eyrich2014,PhysRevLett.99.217208,coey_2010,doi:10.1063/1.3679433,doi:10.1063/1.372577}. For each simulation, we initialized two domain walls of square shape with width $5$~\si{nm} in the $x$ direction with orientation $(m_\mathrm{x},m_\mathrm{y},m_\mathrm{z})=(1/\sqrt{2},-1/\sqrt{2},0)$ after which we minimized (using default settings) this state to find the equilibrium configuration (each wall belonging to a different Bloch state). For the implementation of the RKKY interaction, we use the method proposed in Ref.~\onlinecite{De_Clercq_2017}, where we add custom magnetic fields across non-magnetic spacer layers to act as RKKY fields. We have also verified that a spacer layer of $0.5$ and $1.5$~\si{nm} did not affect the behavior appreciably. 

For Fig. 3 of the main paper and \figref{fig:figure4}, the used cell sizes are $1\times 8 \times 1$~nm (x, y, z) with periodic boundary conditions in the $x$ and $y$ direction equal to 32 repeats. We simulate $6$ magnetic layers ($A=12$~\si{pJ.m^{-1}}) with a thickness of $1$~\si{nm} separated by $1$~\si{nm} thick spacers. Once again, for each simulation, we initialized two domain walls of square shape with width $5$~\si{nm} in the $x$ direction with orientation $(m_\mathrm{x},m_\mathrm{y},m_\mathrm{z})=(1/\sqrt{2},-1/\sqrt{2},0)$ after which we minimized (using default settings) this state to find the equilibrium configuration. We swept the iDMI $D$ and interlayer coupling $J$ (implemented as defined above) for different combinations of $M_S$ and $K_\mathrm{eff}$ to produce the phase diagrams as shown in \figref{fig:figure4} and Fig. 3 of the main text.

Lastly, for \figref{fig:figure5}b, where we show the domain periodicity as a function of $J$, we simulated two $2$~\si{nm} thick magnetic layers (with $A=9$~\si{pJ.m^{-1}}, cell size in the $z$ direction equal to $1$~\si{nm} and the rest as defined in the first paragraph) separated by a $1$~\si{nm} thick spacer. We then varied the size of the simulation box in the $x$ direction and determined for which periodicity there was a minimum in the energy density.

\section{Supporting measurements}

\begin{figure}
\centering
\includegraphics[width=\textwidth]{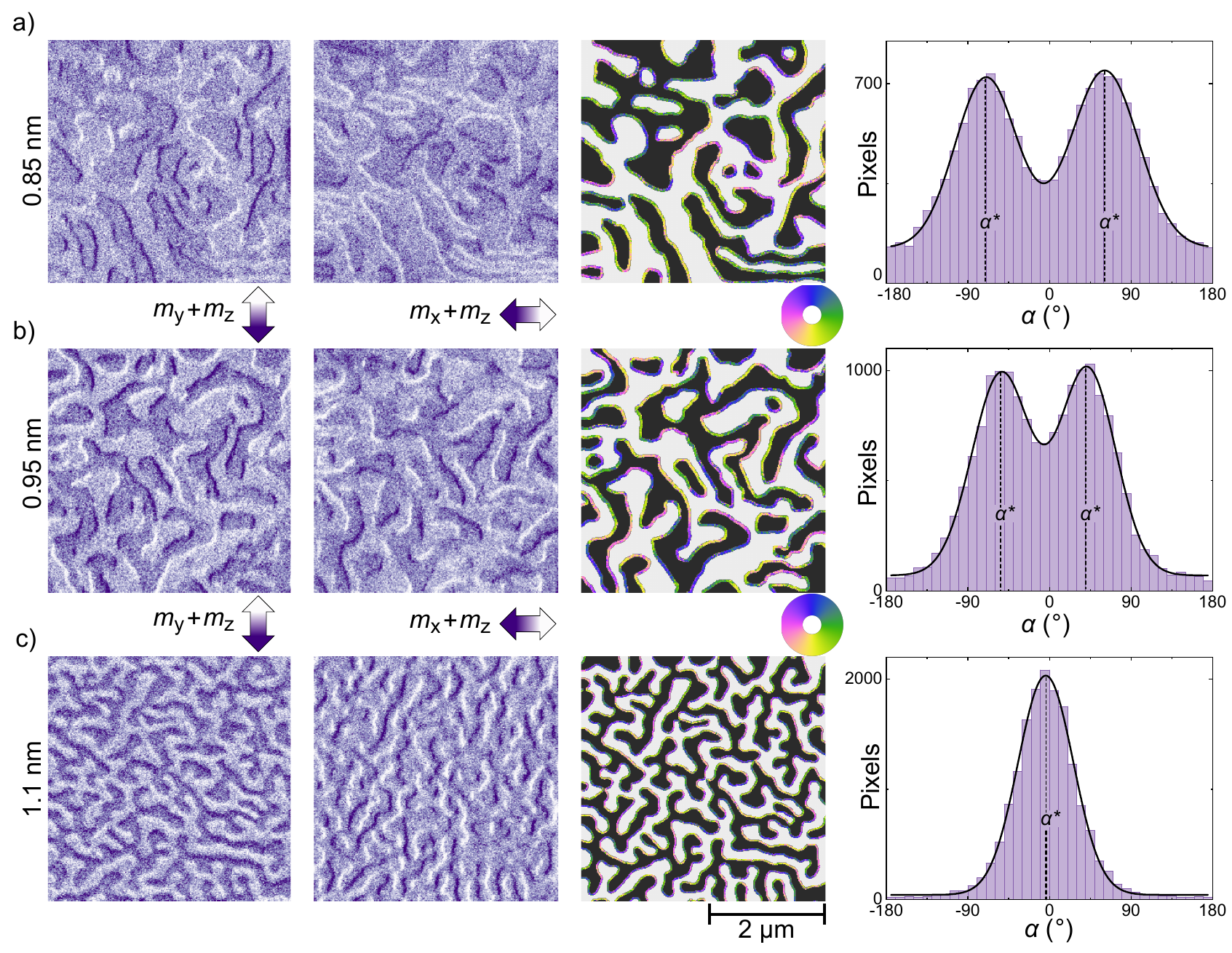}
\caption{\label{fig:figure1}a) SEMPA images of the top magnetic layer at an Ir spacer thickness of $t=0.85$~\si{nm}. The first two images show $m_\mathrm{y}$ + out-of-plane contrast ($m_\mathrm{z}$) and $m_\mathrm{x}$ + out-of-plane contrast ($m_\mathrm{z}$) for the same area. The in-plane magnetic contrast direction is indicated by the arrow underneath the images. In the third image the composite image is depicted, where the in-plane magnetization is indicated by the color wheel and the out-of-plane contrast by the white and black areas (up and down magnetization, respectively). The fourth image is the histogram of $\alpha$ for all pixels in the domain wall of the composite image. A double Gaussian fit (solid line) with maximums $\alpha^\ast$ are depicted. b) Same series of images as in a) for an Ir spacer thickness of $t=0.95$~\si{nm} and in c) for $t=1.1$~\si{nm}. The solid line is a Gaussian fit. The scale bar on the bottom right holds for all images.}
\end{figure}

In this section we present additional SEMPA measurements to support the findings of the main paper. The SEMPA measurements resulting in the composite image and the histogram of Fig. 2d and f in the main paper can be found in \figref{fig:figure1}a. The first image shows the magnetization contrast in $m_\mathrm{y}$ and the second image $m_\mathrm{x}$, as indicated by the arrows underneath the image. A slight out-of-plane contrast ($m_\mathrm{z}$) is observed in both images. The third image is the composite image as depicted in the main text (Fig. 2d). For every pixel in the domain wall an angle $\alpha$ is assigned (see main text for details) and this results in the histogram on the right, which is the same histogram as depicted in Fig. 2f of the main text. The solid line is a double Gaussian fit. In \figref{fig:figure1}b and c the same series of measurements and analysis is shown for an Ir spacer layer thickness of $t=0.95$~\si{nm} and $t=1.1$~\si{nm}, respectively.

\begin{figure}
\centering
\includegraphics[width=0.5\textwidth]{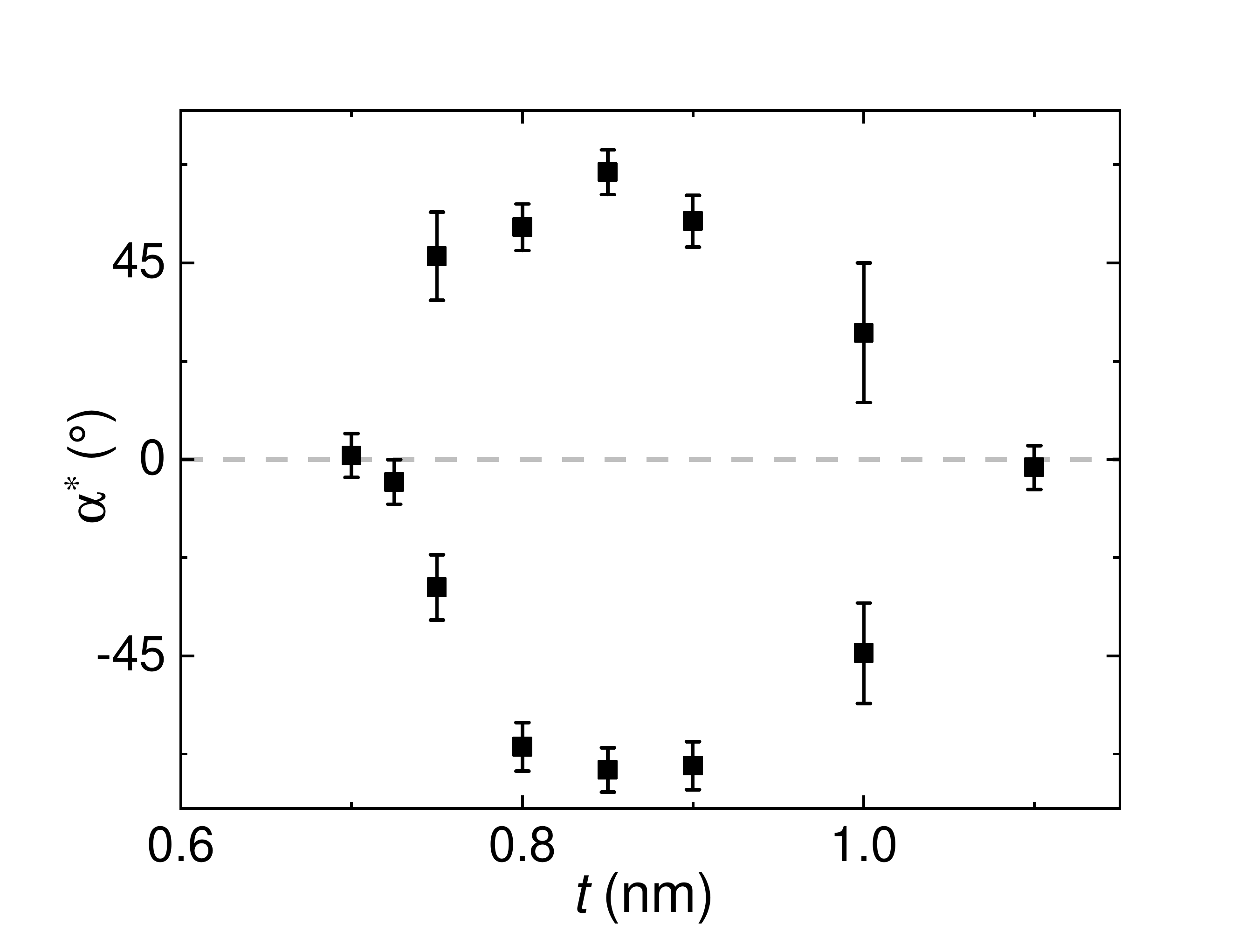}
\caption{\label{fig:figure2} The maximum(s) $\alpha^\ast$ of the (double) Gaussian fits of the histograms as a function of the Ir thickness $t$ on a nominally identical sample to the one presented in the main paper. }
\end{figure}

A second series of measurements is conducted on a nominally identical sample as studied in the main paper to confirm the reproducibility. In \figref{fig:figure2} the results of $\alpha^\ast$ (maximum(s) of the (double) Gaussian fits of the histograms) as a function of the Ir spacer layer $t$ are presented. The same trend as in the main paper (Fig. 2g) is observed, where degenerate Bloch-N\'eel walls are formed for $t=0.75-1.0$~\si{nm} and where CW N\'eel walls are measured for thinner and thicker Ir layers. A maximum angle $\abs{\alpha^\ast}$ is reached at $t=0.85$~\si{nm} and coincides with the findings of the main paper. 

\section{Antiferromagnetic RKKY interaction} \label{spin-flop}
\begin{figure}
\centering
\includegraphics[width=1\textwidth]{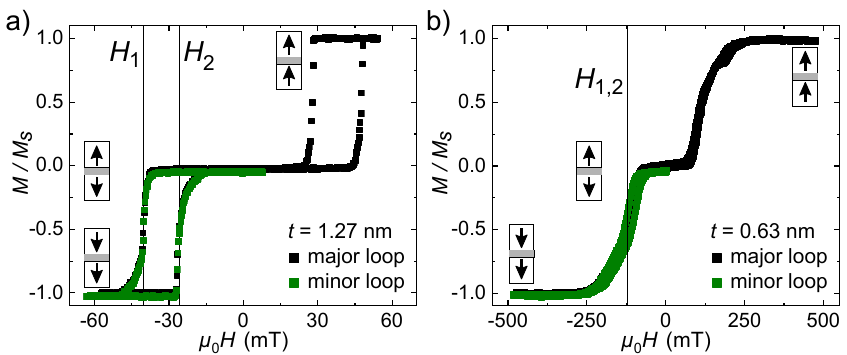}
\caption{\label{fig:figurespin} a) Major and minor hysteresis loop of the antiferromagnetically coupled magnetic bilayer at $t=1.27$~\si{nm} measured by MOKE. The switching fields $H_1$ and $H_2$ are denoted.  b) Major and minor hysteresis loop for an Ir thickness of $t=0.63$~\si{nm}. The insets in a) and b) schematically show the magnetization configuration of the magnetic bilayer. }
\end{figure}

In this section we determine the coupling strength $J$ of the antiferromagnetic RKKY interaction from hysteresis loops. In \figref{fig:figurespin}a a major and minor hysteresis loop,  measured via the magneto-optical Kerr effect (MOKE), are shown for an Ir thickness of $t=1.27$~\si{nm}, corresponding to the second antiferromagnetic region. Three magnetization states are visible in the major loop and they are schematically depicted in the insets. In the field range from $-30$ to $30$~\si{mT} the magnetic layers are coupled antiferromagnetically, but for a large enough positive (negative) applied field the Zeeman energy exceeds the antiferromagnetic coupling $J$ and this results in a parallel magnetization alignment along the field direction. The antiferromagnetic coupling strength of the bilayer can be determined from the switching fields $H_1$ and $H_2$ in the minor loop with the following formula \cite{PhysRevB.50.13505}: $J=1/2\,\mu_0(H_1+H_2)M_Sd$, where $M_S$ is the saturation magnetization of a Co-Ni layer, and $d$ the thickness of that layer.

In \figref{fig:figurespin}b we show a typical hysteresis loop measured by MOKE in the first antiferromagnetic region ($t=0.63$~\si{nm}). Two switches are observed in the major loop, indicating that the magnetic layers are antiferromagnetically coupled and the insets show the magnetization alignment. Sharp switches as measured in \figref{fig:figurespin}a are absent however. We therefore assign the switching fields $H_1$ and $H_2$ at the field where $M=-0.5 M_S$ in the minor loop. The coupling strength $J$ is calculated from these switching fields as mentioned above.  
   
\section{Dependence of the uniform chirality on $M_S$ and $K_\mathrm{eff}$}

\begin{figure}
\centering
\includegraphics[width=0.5\textwidth]{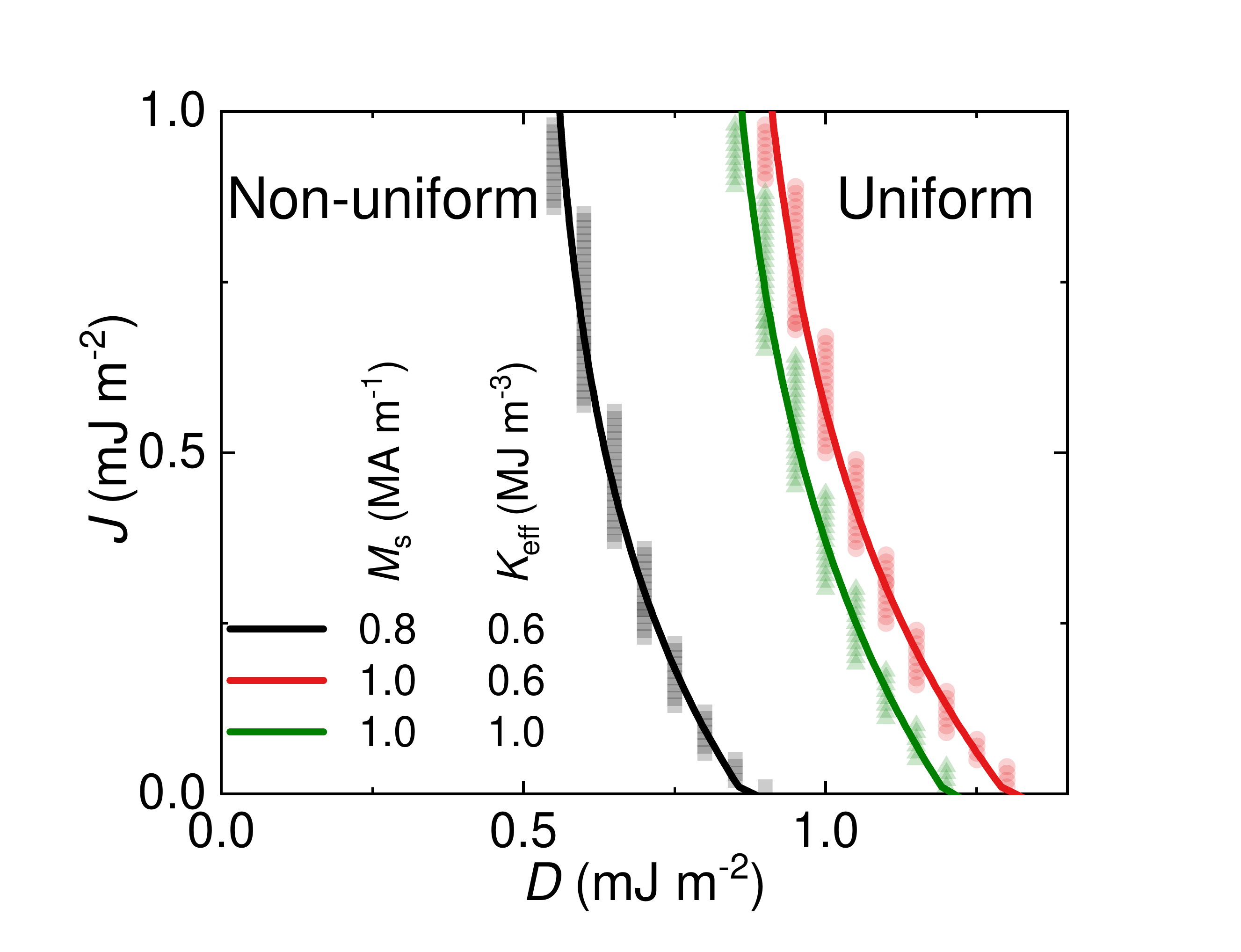}
\caption{\label{fig:figure4} Transition lines separating the non-uniform and uniform magnetization across the magnetic multi-layer for different values of $M_S$ and $K_\mathrm{eff}$. The black curve is the transition line depicted in Fig. 3b in the main paper. }
\end{figure}

Here we discuss the dependence of the results presented in Fig. 3b of the main paper on different values of the saturation magnetization $M_S$ and the effective anisotropy $K_\mathrm{eff}$. In the main paper the micromagnetic simulations are shown for $M_S=0.8$~\si{MA.m^{-1}} and $K_\mathrm{eff}=0.6$~\si{MJ.m^{-3}} and the transition line from Fig. 3b is plotted in black in \figref{fig:figure4}. This dashed line separates the non-uniform magnetization region (left side) from the uniform magnetic chirality region (right side).  

Increasing $M_S$ will shift the transition line towards higher $D$ values (red line), which is equivalent to adding more magnetic volume to the system and thereby increasing the effects of the dipolar field. In that case a larger $D$ is needed to obtain a uniform magnetic chirality. 

To observe the effect of $K_\mathrm{eff}$ we compare the red and green curve and from \figref{fig:figure4} we find that increasing $K_\mathrm{eff}$ results in a shift of the transition line towards lower $D$ values. This dependence is also found in Ref. \cite{PhysRevB.95.174423} in the absence of an RKKY interaction. Apart from the shift in $D$ the functional dependence of the transition line only shows minor changes for different values of $M_S$ and $K_\mathrm{eff}$. 

\section{Dependence of the domain size on $J$} \label{domain_size}

\begin{figure}
\centering
\includegraphics[width=\textwidth]{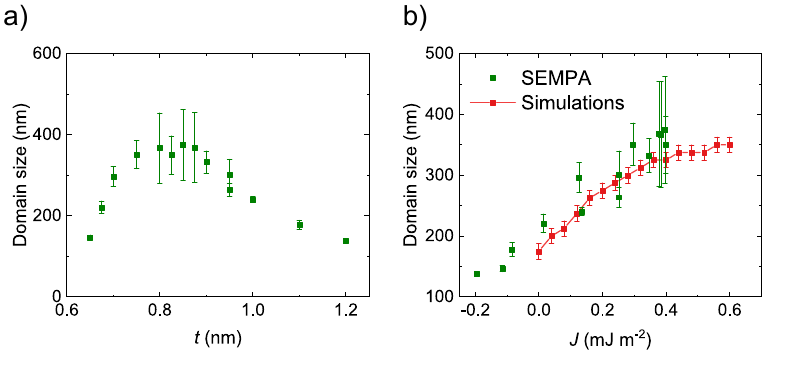}
\caption{\label{fig:figure5} a) Average domain size of the out-of-plane domains from the composite image as a function of the Ir spacer layer thickness. b) Domain size as a function of $J$. The thicknesses $t$ of the measurements in a) are converted to a coupling strength $J$ using a similar fit as shown in Fig. 2h.}
\end{figure}

In this section we discuss the experimentally and theoretically obtained domain sizes as a function of the ferromagnetic RKKY coupling. Comparing the SEMPA images from Fig. 2c and d in the main paper and \figref{fig:figure1}, we find that the average out-of-plane domain size for the as-deposited samples varies strongly with different Ir thicknesses. In \figref{fig:figure5}a we plotted the extracted average domain size for different thicknesses of the Ir spacer layer $t$. The domain sizes are extracted using a spatial FFT analysis on the domains of the composite images (all images are $4\times4$~\si{{\mu m}^2}).

As can be seen from \figref{fig:figure5}a, the domain size increases as a function of $t$, until a maximum is reached between $t=0.8$ and $t=0.9$~\si{nm}. Afterwards a decrease in the domain size is observed and similar domain sizes are obtained for $t=0.65$~\si{nm} and $t>1.1$~\si{nm}. The behavior seems to coincide with the oscillatory behavior of the ferromagnetic RKKY coupling as discussed in the main paper. In \figref{fig:figure5}b the domain sizes from \figref{fig:figure5}a are plotted in green as a function of $J$. Here, we used $A=9$~\si{pJ.m^{-1}} and obtained a similar fit as shown in Fig. 2h in the main paper to convert the Ir thickness $t$ to a value for the coupling strength $J$. Additionally, we plotted in red the domain sizes extracted from micromagnetic simulations as a function of $J$. We had to use $A=9$~\si{pJ.m^{-1}} to obtain similar domain sizes in the simulations and experiments for $J=0$~\si{mJ.m^{-2}}, but the iDMI from the (bottom) Pt/Co interface might have a similar effect \cite{Han2016}. 

In both cases an increase of the domain size as a function of $J$ is observed. Moreover, we find from the simulations that the domain size saturates for larger $J$ values, which coincides with the saturation of $\alpha_\mathrm{top}$ as depicted in Fig. 1c of the main paper. These findings indicate that the ferromagnetic RKKY interaction influences the magnetization inside the domain wall and thereby increases the domain wall energy. The domain size is determined by a competition between the domain wall energy and dipolar energy and since the latter remains constant, the domains grow in size \cite{draaisma1987magnetization}. As soon as (nearly) Bloch walls are formed, the domain wall energy and thus the domain size stabilizes. 

In literature various domain wall models are used to determine magnetic parameters, like the iDMI, from demagnetized domain patterns \cite{draaisma1987magnetization,HELLWIG200713,PhysRevB.95.174423,PhysRevB.96.144408}. The data in \figref{fig:figure5}b suggests that changes in the domain wall energy (and thereby the domain pattern) can be caused by any interaction acting on the magnetization in the domain wall. All these contributions should therefore be considered to be able to accurately extract magnetic parameters from domain patterns. 

\bibliography{references}